\documentclass{sig-alternate}
\usepackage{color}
\usepackage{colortbl}
\usepackage{tabularx}

\usepackage{times} 

\usepackage[
bookmarks=true%
,bookmarksnumbered=true%
,hypertexnames=false%
,breaklinks=true%
]{hyperref} 
\hypersetup{
  pdfauthor = {Vercoustre, Thom, Pehcevski},
  pdftitle = {Entity ranking in Wikipedia},
  pdfsubject = {entity ranking},
  pdfkeywords = {H.3 [Information Storage and Retrieval]},
  pdfcreator = {LaTeX with hyperref package},
  pdfproducer = {pdflatex}
}

\usepackage[square,comma,numbers,sort&compress,sectionbib]{natbib}
\usepackage{latexsym}

\newenvironment{itemize*}%
  {\begin{itemize}%
    \setlength{\itemsep}{0pt}%
    \setlength{\parskip}{0pt}}%
  {\end{itemize}}

\newbox\itembox
\def\itemlistlabel#1{#1\hfill}
\def\itemlist#1{\setbox\itembox=\hbox{#1}%
  \list{}{\labelwidth\wd\itembox
    \leftmargin\labelwidth
    \advance\leftmargin by\itemindent
    \advance\leftmargin by\labelsep
    \let\makelabel\itemlistlabel}}

\toappear{\the\boilerplate\par
{\confname{\the\conf}} \the\confinfo\par \the\copyrightetc.}

\conferenceinfo{SAC'08}{March 16-20, 2008, Fortaleza, Cear\'{a}, Brazil}
\CopyrightYear{2008} 
\crdata{978-1-59593-753-7/08/0003}

\title{Entity Ranking in Wikipedia}

\numberofauthors{3} 
\author{
\alignauthor
Anne-Marie Vercoustre\\
       \affaddr{INRIA}\\
       \affaddr{Rocquencourt, France}\\
       \email{\normalsize anne-marie.vercoustre@inria.fr}
\and
\alignauthor
James A. Thom\\
       \affaddr{RMIT University}\\
       \affaddr{Melbourne, Australia}\\
       \email{\normalsize james.thom@rmit.edu.au}
\and
Jovan Pehcevski\\
       \affaddr{INRIA}\\
       \affaddr{Rocquencourt, France}\\
       \email{\normalsize jovan.pehcevski@inria.fr}
}

\begin{document}

\maketitle

\begin{abstract}
  
The traditional entity extraction problem lies in the ability of extracting named entities from plain
text using natural language processing techniques and intensive
training from large document collections.
Examples of named entities include organisations, people, locations, or dates. 
There are many research activities involving named entities;
we are interested in entity ranking in the field of information retrieval.
In this paper, we describe our approach to identifying and ranking
entities from the INEX Wikipedia document collection.
Wikipedia offers a number of interesting features for entity identification and ranking that we first introduce.
We then describe the principles and the architecture of our entity ranking system,
and introduce our methodology for evaluation.
Our preliminary results show that the use of categories and the link structure of Wikipedia, together with entity examples, can significantly improve retrieval effectiveness.

\end{abstract}
\vspace{1mm}

{\small
\noindent
{\bf Categories and Subject Descriptors:} 
H.3 {[Information Storage and Retrieval]}: {H.3.3 Information Search
and Retrieval}\\
\vspace{1mm}
}

\section{Introduction}
\label{sec:int}

Information systems contain references to many named entities.
In a well-structured database system it is exactly clear what are references to named entities,
whereas in semi-structured information sources (such as web pages) it is harder to identify them within a text.
An entity could be, for example, an organisation, a person, a location, or a date. 
Because of the importance of named entities, several very active and related research areas have emerged in recent years,
including: entity extraction/tagging from texts,
entity reference solving,
entity disambiguation,
question-answering, 
expert search, and entity ranking (also known as entity retrieval). 

The traditional entity extraction problem is to extract named entities from plain text using natural language processing techniques 
or statistical methods and intensive training from large collections~\cite{seki:prot04,cunn:acl02}. 
Benchmarks for evaluation of entity extraction have been performed for the Message Understanding Conference (MUC)~\cite{Sund:muc91} and for the Automatic Content Extraction (ACE) program~\cite{ace:2006}. 
Training is done on a large number of examples in order to identify extraction patterns (rules).
The goal is to eventually tag those entities and use the tag names to support future information retrieval. 
\newpage
In the context of large collections such as the web or Wikipedia, it is not possible, nor even desirable,
to tag in advance all the entities in the collection,
although many occurrences of named entities in the text may be used as anchor text for sources of hypertext links.
Instead, since we are dealing with semi-structured documents (HTML or XML), we could exploit the explicit document structure to infer effective extraction patterns or algorithms.

The goal of {\em entity ranking} is to retrieve entities as answers to a query.
The objective is not to tag the names of the entities in documents but rather to get back a list of the relevant entity names
(possibly each entity with an associated description).
For example, the query ``European countries where I can pay
with Euros''~\cite{vries:ent06} should return a list of entities (or pages) representing relevant
countries, and not a list of entities about the Euro and similar
currencies.

The Initiative for the Evaluation of XML retrieval (INEX) has a new track on entity ranking~\cite{vries:ent07}, using Wikipedia as its document collection.
This track proposes two tasks:
a task where the category of the expected entity answers is provided; and
a task where a few (two or three) examples of the expected entity answers are provided.
The inclusion of target categories (in the first task) and example entities (in the second task)
makes these quite different tasks from full-text retrieval.
The combination of the query and example entities (in the second task) makes it quite different from an application such as Google Sets\footnote{http://labs.google.com/sets} where only entity examples are provided.

In this paper, we identify some important principles for entity ranking that we incorporate into an
architecture which allows us to tune, evaluate, and improve our approach as it develops.
Our entity ranking approach is based on three ideas: 
(1) using full-text similarity with the query, 
(2) using popular links (from highly scored pages), and 
(3) using category similarity with the entity examples.



\section{Related work}
\label{sec:related}

Our entity ranking approach gets its inspiration from wrapping technology, entity extraction, 
the use of ontologies for entity extraction or entity disambiguation, and link analysis.

\subsection*{Wrappers}

A wrapper is a tool that extracts information (entities or values) from a document, or a set of documents, with a purpose of reusing information in another system.
A lot of research has been carried out in this field by the database community,
mostly in relation to querying heterogeneous databases~\cite{verc:oois97,sahu:vldb99,adel:sigmod99,kush:ai00}.
More recently, wrappers have also been built to extract information from web pages with different applications in mind,
such as product comparison, reuse of information in virtual documents, or building experimental data sets.
Most web wrappers are either based on
scripting languages~\cite{verc:oois97,sahu:vldb99} that are very close to current XML query languages, or use wrapper induction~\cite{adel:sigmod99,kush:ai00} that learn rules for extracting information.

%

To prevent wrappers breaking over time without notice when pages change, Lerman et al.~\cite{lerm:jair03} propose using machine learning for wrapper verification and re-induction.
Rather than repairing a wrapper over changes in the web data, Callan and Mitamura~\cite{callan:2002} propose generating the wrapper dynamically~---~that is at the time of wrapping, using data previously extracted and stored in a database.
The extraction rules are based on heuristics around a few pre-defined lexico-syntactic HTML patterns such as lists, tables, and links.
The patterns are weighted according to the number of examples they recognise; the best patterns are used to dynamically extract new data.

Our system for entity ranking also works dynamically, at query time instead of at wrapping time.
We also use weighting algorithms based
on links
that are well represented in web-based collections, as well as knowledge of categories, a specific Wikipedia feature.

\subsection*{Entity extraction}

 
Recent research in named entity extraction has developed approaches that are not language dependant and do not require lots of linguistic knowledge.
McNamee and Mayfield~\cite{mcna:coling02} developed a system for entity extraction based on training on a large set of very low level textual patterns found in tokens.
Their main objective was to identify entities in multilingual texts and classify them into one of four classes (location, person, organisation, or ``others''). Cucerzan and Yarowsky~\cite{cuce:emnlp99} describe
and evaluate a language-independent bootstrapping algorithm based on iterative learning and re-estimation of contextual and morphological patterns. It achieves competitive performance when trained on a very short labelled name list.

\subsection*{Using ontology for entity extraction}

Other approaches for entity extraction are based on the use of external resources, such as an ontology or a dictionary.
Popov et al.~\cite{popo:iswc03} use a populated ontology for entity extraction, while Cohen and Sarawagi~\cite{cohe:kdd04} exploit a dictionary for named entity extraction. 
Tenier et al.~\cite{teni:egc06} use an ontology for automatic semantic annotation of web pages. Their system firstly identifies the syntactic structure that characterises an entity in a page, and then uses subsumption to identify the more specific concept to be associated with this entity.

\subsection*{Using ontology for entity disambiguation}

Hassell et al.~\cite{hass:iswc06} use a ``populated ontology'' to assist in disambiguation of entities,
such as names of authors using their published papers or domain of interest. 
They use text proximity between entities to disambiguate names
(e.g. organisation name would be close to author's name).
They also use text co-occurrence, for example for topics relevant to an author.
So their algorithm is tuned for their actual ontology,
while our algorithm is more based on the the categories and the structural properties of the Wikipedia.

Cucerzan~\cite{cuce:emnlp07} uses Wikipedia data for named entity disambiguation.
He first pre-processed a version of the Wikipedia collection (September 2006),
and extracted more than 1.4 millions entities with an average of 2.4 surface forms by entities.
He also extracted more than one million (entities, category) pairs that were further filtered down to 540 thousand pairs.
Lexico-syntactic patterns, such as titles, links, paragraphs and lists, are used to build co-references of entities
in limited contexts.
The knowledge extracted from Wikipedia is then used for improving entity disambiguation in the context of web and news search. 

\subsection*{Link Analysis (PageRank and HITS)}

Most information retrieval (IR) systems use statistical information concerning the distribution of the query terms
to calculate the query-document similarity.
However, when dealing with
hyperlinked environments such as the web or Wikipedia, link analysis is also important.
PageRank and HITS are two of the most popular algorithms that use link analysis to improve web search performance.

PageRank, an algorithm proposed by Brin and Page~\cite{Google}, is a link analysis algorithm that assigns a numerical weighting to each page of a hyperlinked set of web pages. The idea of PageRank is that a web page is a good page if it is popular, 
that is if many other (also preferably popular) web pages are referring to it.

In HITS (Hyperlink Induced Topic Search), {\it hubs} are considered to be web pages that have links pointing to many {\it authority} pages~\cite{Klein99}. However, unlike PageRank where the page scores are calculated independently of the query by using the complete web graph, in HITS the calculation of hub and authority scores is query-dependent; here, the so-called {\em neighbourhood graph} includes not only the set of top-ranked pages for the query, but it also includes the set of pages that either point to or are pointed to by these pages.

We use the idea behind PageRank and HITS in our system; however, instead of counting every possible link referring to an entity page in the collection (as with PageRank), or building a neighbourhood graph (as with HITS), we only consider pages that are pointed to by a selected number of top-ranked pages for the query. This makes our link ranking algorithm query-dependent (just like HITS), allowing it to be dynamically calculated at query time.


\section{INEX Entity Ranking track}
\label{sec:INEXertrack}

The INEX Entity ranking track was proposed as a new track in 2006, but will only start in 2007.
It will use the Wikipedia XML document collection (described in the next section) that has been used by various INEX tracks in 2006~\cite{INEX06-Overview}.
Two tasks are planned for the INEX Entity ranking track in 2007~\cite{vries:ent07}:

\begin{description}
\item[task1:] entity ranking, where the aim is to retrieve entities of a given category satisfying a topic described by a few query terms;
\item[task2:] list completion, where given a topic text and a number of examples, the aim is to complete this partial list of answers. 
\end{description}

\begin{figure}
\small
\hrule
\begin{verbatim}

<inex_topic> 
<title>
European countries where I can pay with Euros
</title>
<description>
I want a list of European countries where
I can pay with Euros.
</description>
<narrative>
Each answer should be the article about a specific
European country that uses the Euro as currency.
</narrative>
<entities>
   <entity id="10581">France</entity> 
   <entity id="11867">Germany</entity>
   <entity id="26667">Spain</entity>
</entities>
<categories>
   <category id="61">countries<category>
</categories>
</inex_topic>
\end{verbatim}
\vskip -6pt
\hrule
\caption{Example INEX 2007 entity ranking topic}
\label{fig:topic}
\end{figure}

Figure~\ref{fig:topic} shows an example INEX entity ranking topic;
the {\tt title} field contains the query terms, 
the {\tt description} provides a natural language summary of the information need,
and the {\tt narrative} provides an explanation of what makes an entity answer relevant.
In addition, the {\tt entities} field provides a few of the expected entity answers for the topic (task 2),
while the {\tt categories} field provides the category of the expected entity answers (task 1).

\section{The INEX Wikipedia corpus}
\label{sec:wikipedia}


Wikipedia is a well known web-based, multilingual, free content encyclopedia written collaboratively by contributors from around the world.
As it is fast growing and evolving it is not possible to use the actual online Wikipedia for experiments.
Denoyer and Gallinari~\cite{Wikipedia} have developed an XML-based corpus founded on a snapshot of the Wikipedia, which has been used by various INEX tracks in 2006.
It differs from the real Wikipedia in some respects (size, document format, category tables), but it is a very realistic approximation.
Specifically, the INEX Wikipedia XML document corpus retains the main characteristics of the online version, although they have been implemented through XML tags instead of the initial HTML tags and the native Wikipedia structure.
The corpus is composed of 8 main collections, corresponding to 8 different languages.
The INEX 2007 Entity ranking track will use the 4.6 Gigabyte English sub-collection which contains 659,388 articles.




\subsection{Entities in Wikipedia}

In Wikipedia, an entity is generally associated with an article (a Wikipedia page) describing this entity.
Nearly everything can be seen as an entity with an associated page, including
countries, famous people, organisations, places to visit, and so forth.

The entities have a name (the name of the corresponding page) and a unique ID in the collection.
When mentioning such an entity in a new Wikipedia article, authors are encouraged to link at least the first occurrence of the entity name to the page describing this entity.
This is an important feature as it allows to easily locate potential entities, which is a major issue in entity extraction from plain text. 
Consider the following extract from the Euro page.
\begin{quote}
``The {\bf euro} \ldots
is the official \underline{currency} of the \underline{Eurozone} (also known as the Euro Area), which consists of the \underline{European} states of \underline{Austria}, \underline{Belgium}, \underline{Finland}, \underline{France}, \underline{Germany}, \underline{Greece}, \underline{Ireland}, \underline{Italy}, \underline{Luxembourg}, \underline{the Netherlands}, \underline{Portugal}, \underline{Slovenia} and \underline{Spain}, and will extend to include \underline{Cyprus} and \underline{Malta} from 1 January 2008.''
\end{quote}
All the underlined words (hypertext links that are usually highlighted in another colour by the browser) can be seen as occurrences of entities that are each linked to their corresponding pages.
In this extract, there are 18 entity references of which 15 are country names; these countries are all ``European Union member states'',
which brings us to the notion of category in Wikipedia.

\subsection{Categories in Wikipedia}

Wikipedia also offers categories that authors can associate with Wikipedia pages.
New categories can also be created by authors, although they have to follow Wikipedia recommendations in both creating new categories and associating them with pages.  
For example, the Spain page is associated with the following categories: ``Spain'', ``European Union member states'', ``Spanish-speaking countries'', ``Constitutional monarchies'' (and some other Wikipedia administrative categories).
There are 113,483 categories in the INEX Wikipedia XML collection, which are organised in a graph of categories.
Each page can be associated with many categories (2.28 as an average). 
Some properties of Wikipedia categories (explored in more detail by Yu et al.~\cite{yu:cikm07}) include:
\begin{itemize*}
\item	a category may have many sub-categories and parent categories;
\item	some categories have many associated pages (i.e. large extension), while others have smaller extension;
\item	a page that belongs to a given category extension generally does not belong to its ancestors' extension;
\item	the sub-category relation is not always a subsumption relationship; and
\item	there are cycles in the category graph.
\end{itemize*}

When searching for entities it is natural to take advantage of the Wikipedia categories since they give hints on whether the retrieved entities are of the expected type.
For example, if looking for `author' entities, pages associated with the category ``novelists'' may be more relevant than pages associated with the category ``books''.

\section{Our approach}
\label{sec:approach}

In this work, we are addressing the task of ranking entities in answer to a query
supplied with a few examples (task 2).
Our approach is based on the following principles for entity answer pages. A good page:
\begin{itemize*}
\item answers the query (or a query extended with the examples),

\item is associated with a category close to the categories of the entity examples (we use a similarity function between the categories of a page and the categories of the examples),

\item is pointed to by a page answering the query (this is
an adaptation of the HITS~\cite{Klein99} algorithm to the problem of entity
ranking; we refer to it as a linkrank algorithm), and

\item is pointed to by contexts with many occurrences of the entity examples. We currently use the full page as the context 
when calculating the scores in our linkrank algorithm. Smaller contexts such as paragraphs, lists, or tables 
have been used successfully by others~\cite{liu:kdd03}. 

\end{itemize*}

We have built a system based on the above principles,
where candidate pages
are ranked by combining three different scores: a linkrank score, a
category score, and the initial search engine similarity score.
We use Zettair,\footnote{http://www.seg.rmit.edu.au/zettair/} 
a full-text search engine developed by RMIT University, which returns pages ranked by their similarity score to the query. 
We use the Okapi BM25 similarity measure as it was effective on the INEX Wikipedia collection~\cite{Iskandar-INEX06}.

Our system involves several modules for
processing a query, submitting it to the search engine, applying our entity
ranking algorithms, and finally returning a ranked list of entities, including: 
\begin{description}

\item[the topic module] takes an INEX topic as input (as the topic example shown in Figure~\ref{fig:topic})
and generates the corresponding Zettair query and the list of entity examples (as an option, the example
entities may be added to the query);

\item[the search module] sends the query to Zettair and returns a list of ranked Wikipedia pages (typically 1500); and

\item[the link extraction module] extracts the links from a selected number of highly ranked pages together with the XML paths of the links
(we discard external links and internal collection links that do not refer to existing pages in the collection).
\end{description}
Using the pages found by these modules, we calculate
a global score for each page (see~\ref{global}) as a linear combination of the normalised scores coming out from the following three functions:

\begin{itemize}

\item {\bf the linkrank function}, which calculates a weight for a page based (among other things) on the number of links to this page (see~\ref{linkrank});

\item {\bf the category similarity function}, which calculates a weight for a page based on the similarity of the page categories with those of the entity examples (see~\ref{category}); and

\item {\bf the full-text IR function}, which calculates a weight for a page based on its initial Zettair score (see~\ref{zet}).

\end{itemize}

\begin{figure}
\centering
\epsfig{file=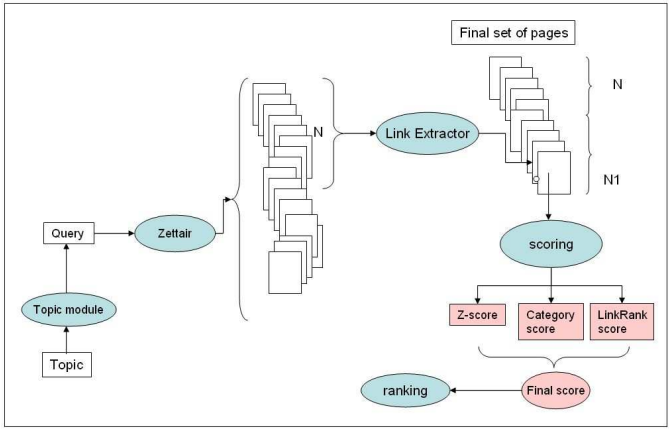, width=3.5in}
\caption{Process for Entity ranking}
\label{process}
\end{figure}

The overall process for entity ranking is shown in Figure~\ref{process}.
The architecture provides a general framework for evaluating entity ranking
which allows for replacing some modules by more advanced modules, 
or by providing a more efficient implementation of a module.
It also uses an evaluation module (not shown in the figure) to assist in tuning the system by
varying the parameters and to globally evaluate the entity ranking approach.



\subsection{LinkRank score}
\label{linkrank}

The linkrank function calculates a score for a page, based on the number of
links to this page, from the first N pages returned by the search engine in
response to the query.
We carried out some experiments with different
values of N and found that N=20 was an acceptable compromise between performance
and discovering more potentially good entities.
The linkrank function can be implemented in a variety of ways: by weighting
pages that have more links referring to them from 
higher ranked pages (the initial N pages), or from pages containing larger number of entity examples, or a combination of the two. 
We have implemented a very basic linkrank function that, for a target entity page $t$, takes into 
account the Zettair score of the referring page $z(p_r)$, the number of distinct entity examples in the referring page $\#ent(p_r)$, and the number of reference links to the target page $\#links(p_r,t)$:
\begin{equation}\displaystyle
	S_L(t) = \sum_{r=1}^N z(p_r) \cdot g(\#ent(p_r)) \cdot f(\#links(p_r,t))
\end{equation}

\noindent where $g(x) = x + 0.5$ (we use 0.5 to allow for cases where there are no entity examples in the referring page) and $f(x) = x$ (as there is at least one reference link to the target page).

\subsection{Category similarity score}
\label{category}

There has been a lot of research on similarity between concepts of
two ontologies, especially for addressing the problem of mapping or updating ontologies~\cite{blan:interop05}. 
Similarity measures between concepts of the same ontology cannot be
applied directly to Wikipedia categories, mostly because the notion of
sub-categories in Wikipedia is not a subsumption relationship. Another
reason is that categories in Wikipedia do not form a hierarchy (or a set
of hierarchies) but a graph with potential cycles. Therefore tree-based
similarities~\cite{blan:ifcs06} either cannot be used or their applicability is limited.

However, the notions of ancestors, common ancestors, and shorter paths between categories can still be used, 
which may allow us to define a distance between
the set of categories associated with a given page, and the set of categories associated with the entity examples. 
We use a very basic similarity function that is the ratio of common categories between the set of categories associated with the target page $\mathsf{cat}(t)$ and the union of the categories associated with the entity examples $\mathsf{cat}(E)$:
\begin{equation}\displaystyle
	S_C(t) = \frac{| \mathsf{cat}(t) \cap \mathsf{cat}(E)|}{ | \mathsf{cat}(E) | }
\end{equation}


\subsection{Z score}
\label{zet}

The Z score assigns the initial Zettair score to a target page. If the target page does not appear in the list of 1500 ranked pages returned by Zettair, then its Z score is zero:
\begin{equation}
	S_Z(t) = \begin{cases} 
	z(t) & \text{if page }t \text{ was returned by Zettair} \cr
&\cr
0 & \text{otherwise } \cr
\end{cases}
\end{equation}

\subsection{Global score}
\label{global}

The global score $S(t)$ for a target entity page is calculated as a linear combination of three normalised scores, the linkrank score $S_L(t)$, the category similarity score $S_C(t)$, and the Z score $S_Z(t)$:
\begin{equation}
	S(t) = \alpha S_L(t)  + \beta S_C(t) + (1 -  \alpha - \beta) S_Z(t)
\end{equation}
where $\alpha$ and $\beta$ are parameters that can be tuned.
Some special cases let us evaluate the effectiveness of each module in our system:
where only the linkrank score is used
($\alpha=1$, $\beta=0$);
where only the category score is used
($\alpha=0$, $\beta=1$);
and
where only the Z score is used\footnote{This is not the same as the plain Zettair score, as apart from the highest N pages returned by Zettair, the remaining N1 entity answers are all generated by extracting links from these pages.}
($\alpha=0$, $\beta=0$).
More combinations of the two parameters are explored in the training phase of our system.

\begin{table*}[tp]
\caption
{Mean average precision scores for runs using 66 possible $\alpha$--$\beta$ combinations, obtained on the 11 INEX 2006 training topics. Queries sent to Zettair include only terms from the topic title (Q). The MAP score of the plain Zettair run is 0.1091. The numbers in italics show the scores obtained for each of the three individual modules. The best performing MAP score is shown in bold.}
\begin{center}
\begin{tabular}{c c c c c c c c c c c c}
\hline \hline
 & \multicolumn{11}{c}{{\bf Beta ($\beta$)}} \\
 \cline{2-12}
 {\bf Alpha ($\alpha$)} & 0.0 & 0.1 & 0.2 & 0.3 & 0.4 & 0.5 & 0.6 & 0.7 & 0.8 & 0.9 & 1.0  \\
\hline \hline
0.0 & {\em 0.1189} & 0.1374 & 0.1688 & 0.1891 & 0.2190 & 0.2158 & 0.2241 & 0.2295 & 0.2424 & 0.2505 & {\em 0.2382} \\
0.1 & 0.1316 & 0.1479 & 0.1917 & 0.2041 & 0.2299 & 0.2377 & 0.2562 & 0.2669 & 0.2707 & 0.2544 & \\
0.2 & 0.1428 & 0.1644 & 0.1897 & 0.2279 & 0.2606 & 0.2655 & 0.2795 & 0.2827 & 0.2602 & & \\
0.3 & 0.1625 & 0.1893 & 0.2058 & 0.2383 & 0.2703 & 0.2766 & {\bf 0.2911} & 0.2631 & & & \\
0.4 & 0.1774 & 0.1993 & 0.2220 & 0.2530 & 0.2724 & 0.2822 & 0.2638 & & & & \\
0.5 & 0.1878 & 0.2075 & 0.2279 & 0.2517 & 0.2762 & 0.2623 & & & & & \\
0.6 & 0.1979 & 0.2153 & 0.2441 & 0.2460 & 0.2497 & & & & & & \\
0.7 & 0.2011 & 0.2187 & 0.2342 & 0.2235 & & & & & & & \\
0.8 & 0.2016 & 0.2073 & 0.2006 & & & & & & & & \\
0.9 & 0.1939 & 0.1843 & & & & & & & & & \\
1.0 & {\em 0.1684} & & & & & & & & & & \\
\hline \hline
\end{tabular}
\end{center}
\label{Q-MAP-training}
\end{table*}

\section{Experimental results}
\label{sec:results}


\subsection{Evaluation Methodology}



There is no existing set of topics with assessments for entity ranking, although such a set will be developed for the INEX entity ranking track in 2007.
So
we developed our own test collection based on a selection of topics from the INEX 2006 ad hoc track.
We chose 27 topics from the INEX 2006 ad hoc track that we considered were of an ``entity ranking'' nature. For each page that had been assessed as containing relevant information, we reassessed whether or not it was an entity answer, that is whether it {\em loosely} belonged to a category of entity we had {\em loosely} identified as being the target of the topic.
We did not require that the answers should strictly belong to a particular category in the Wikipedia.
If there were example entities mentioned in the original topic, then these were usually used as entity examples in the entity topic.
Otherwise, a selected number (typically 2 or 3) of entity examples were chosen somewhat arbitrarily from the relevance assessments.

We used the first 9 of the 27 topics in our training set,
to which we added two more topics created by hand from the original track description
(one of these extra topics is the Euro example in Figure~\ref{fig:topic}). The remaining 18 topics were our test set.


We use MAP (mean average precision) as our primary method of evaluation, but also report some results with other measures 
which are typically used to evaluate the retrieval performance of IR systems~\cite{TREC-book}.
We first remove the the entity examples both from the list of answers returned by each system and from the relevance assessments 
(as the task is to find entities other than the examples provided).
We calculate precision at rank $r$ as follows:
\begin{equation}\displaystyle
P[r] = \frac{\sum_{i=1}^{r} rel(i)}{r}
\end{equation}

\noindent where $rel(i) = 1$ if the $i^{th}$ article in the ranked list was judged as a relevant entity, $0$ otherwise.
Average precision is calculated as the average of $P[r]$ for each relevant entity retrieved
(that is at natural recall levels); if a system does not retrieve a particular relevant entity, then the precision for that entity is assumed to be zero. MAP is the mean value of the average precisions over all the topics in the training (or test) data set.

We also report on several alternative measures: mean of $P[1]$, $P[5]$, $P[10]$ (mean precision at top 1, 5 or 10 entities returned),
mean R-precision (R-precision for a topic is the $P[R]$, where $R$ is the number of entities that have been judged relevant for the topic). 


\subsection{Training data set (11 topics)}

We used the training data set to determine suitable values for the parameters $\alpha$ and $\beta$. 
We varied $\alpha$ from 0 to 1 in steps of 0.1, and for each value of $\alpha$, we varied $\beta$ from 0 to $(1 - \alpha)$ in steps of 0.1. We observe in Table~\ref{Q-MAP-training} that the highest MAP (0.2911) on the 11 topics is achieved for $\alpha = 0.3$ and $\beta = 0.6$.
We also tried training using mean R-precision instead of MAP as our evaluation measure
where we observed somewhat different optimal values for the two parameters: $\alpha = 0.1$ and $\beta = 0.8$. A reason for this is the relatively small number of topics used in the training data set. We would expect the optimal parameter values obtained by MAP and R-precision to converge if many more training topics are used.


On the training data, we also experimented with adding the names of the example entities to the query sent to Zettair.
However this generally performed worse, 
both for the plain Zettair run and the runs using various score combinations, 
and so it would require a more detailed per-topic analysis in order to investigate why this occurs. 
Accordingly, for the test collection we use queries that only include terms from the topic title, 
and consider the plain Zettair run as a baseline while comparing our entity ranking approaches.

\begin{table}[tp]
\caption
{Performance scores for runs obtained with different evaluation measures, using the 18 INEX 2006 test topics. Queries sent to Zettair include only terms from the topic title (Q). The best performing scores are shown in bold.}
\begin{center}
\begin{tabular}{l c c c c c}
\hline \hline
 & \multicolumn{3}{c}{{\bf {\em P[r]}}} & & \\
 \cline{2-4}
 {\bf Run} & 1 & 5 & 10 & {\bf {\em R-prec}} & {\bf {\em MAP}}  \\
\hline \hline
Zettair & 0.2778 & 0.3000 & 0.2722 & 0.2258 & 0.2023 \\
\hline
$\alpha$0.0--$\beta$0.0 & 0.2778 & 0.3000 & 0.2722 & 0.2363 & 0.2042 \\
$\alpha$0.0--$\beta$1.0 & {\bf 0.5556} & 0.4111 & 0.3444 & 0.3496 & 0.3349 \\
$\alpha$1.0--$\beta$0.0 & 0.0556 & 0.1556 & 0.1278 & 0.1152 & 0.1015 \\
\hline
$\alpha$0.3--$\beta$0.6 & 0.5000 & 0.4444 & 0.3667 & 0.3815 & 0.3274 \\
$\alpha$0.1--$\beta$0.8 & {\bf 0.5556} & {\bf 0.5333} & {\bf 0.4222} & {\bf 0.4337} & {\bf 0.3899} \\
\hline \hline
\end{tabular}
\end{center}
\label{Q-test-scores}
\end{table}

\subsection{Test data set (18 topics)}

In these experiments, we designed runs to compare six entity ranking approaches using the 18 topics in the test data set:
\begin{itemize*}
\item full-text retrieval using Zettair (as a baseline)
\item link extraction and re-ranking using the Z score ($S_Z$)
\item link extraction and re-ranking using the category score ($S_C$)
\item link extraction and re-ranking using the linkrank score ($S_L$)
\item link extraction and re-ranking using two global scores:
\begin{itemize*}
\item ($0.3*S_L+0.6*S_C+0.1*S_Z$)
\item ($0.1*S_L+0.8*S_C+0.1*S_Z$)
\end{itemize*}
\end{itemize*}
The results for these six runs are shown in Table~\ref{Q-test-scores}. 
We observe that the best entity ranking approach is the one that places most of the weight on the category score $S_C$ (run $\alpha$0.1--$\beta$0.8). With both MAP and R-precision, this run performs significantly better ($p < 0.05$) than the plain Zettair full-text retrieval run and the other four runs that use various score combinations in the re-ranking. The run that uses the category score in the re-ranking performs the best among the three runs that represent the three individual modules; however, statistically significant performance difference ($p < 0.05$) is only observed when comparing this run to the worst performing run ($\alpha$1.0--$\beta$0.0) which uses only the linkrank score in the re-ranking.

These results show that the global score (the combination of the three individual scores), optimised in a way to give more weight on the category score, 
brings the best value in retrieving the relevant entities for the INEX Wikipedia document collection.

\section{Conclusion and future work}
\label{sec:conclusion}

We have presented our entity ranking approach for the INEX Wikipedia XML document collection which 
is based on exploiting the interesting structural and semantic properties of the collection.
We have shown in our preliminary evaluations that simple use of the categories and the link structure of Wikipedia, together with the entity examples from the topic, can significantly improve the entity ranking performance compared to a full-text retrieval engine. 

Our current implementation uses very simple linkrank and category similarity functions and offers room for improvement.

To improve the linkrank function, we plan to narrow the context around the entity examples. 
We expect relevant entities to frequently co-occur with the example entities in lists. 
The narrower context could be defined either by fixed XML elements (such as paragraphs, lists, or tables) or it could be
determined dynamically. To determine it dynamically, we plan to identify {\em coherent retrieval elements}
adapting earlier work by Pehcevski et al.~\cite{Pehcevski-Kluwer2005} to identify the element contexts that are most likely to contain lists. 

To improve the category similarity function, we plan to take into account the notion of existing sub-categories and parent categories found in Wikipedia.

We will also be participating in the INEX~2007 entity ranking track, which we expect would enable us to test our approach using a larger set of topics and compare it against alternative approaches.


\smallskip\noindent
\textbf{Acknowledgements} %
This work was undertaken while James Thom was visiting INRIA in 2007.

\renewcommand{\bibsection}{\section*{REFERENCES}}
\providecommand{\bibfont}{\small}
\setlength{\bibhang}{1em}
\setlength{\bibsep}{0.5ex} 
\bibliographystyle{abbrvnat}
\bibliography{strings,mybibfile}

\end{document}